\documentstyle[preprint,osa,epsf]{revtex}
\begin{document}

\title{Dispersion-managed soliton in a strong dispersion map limit
\footnote{\bf Submitted to Optics Letters} }

\author{P. M. Lushnikov$^{1,2}$}

\address{$^1$ Theoretical Division, Los Alamos National Laboratory, MS-B284, Los Alamos, New
Mexico, 87545
\\
$^2$ Landau Institute for Theoretical Physics, Kosygin St. 2, Moscow, 117334, Russia
   }


\maketitle

\begin{abstract}

A dispersion-managed optical system with step-wise periodical variation of dispersion is studied
in a strong dispersion map limit in the framework of path-averaged Gabitov-Turitsyn equation. The
soliton solution is obtained by iterating the path-averaged equation analytically and numerically.
An efficient numerical algorithm for obtaining of DM soliton shape is developed. The envelope of
soliton oscillating tails is found to decay exponentially in time while the oscillations are
described by a quadratic law.
\end{abstract}
~~~~~~~ {\it OCIS codes:}  060.2330, 060.5530, 060.4370, 190.5530, 260.2030.
\\

This information is Copyright© 1999 Personal TeX, Inc. All Rights Reserved.

A dispersion-managed \cite{kogelnik1} (DM) optical fiber is designed to create a low (or even zero)
path-averaged dispersion by periodically alternating dispersion sign along an optical line which
dramatically reduces pulse broadening. Recently dispersion management has become an essential
technology for development of ultrafast high-bit-rate optical communication lines
\cite{nakazawa1,smithknox1,gabtur1,kumar1,kaup1,mamyshev1,turitsyn1}. Lossless propagation of
optical pulse in DM fiber is described by a nonlinear Schr\"odinger equation (NLS) with
periodically varying dispersion $d(z)$:
\begin{equation}
i u_z +   d(z) u_{tt} +  |u|^{2} u =0,  \label{nls1}
\end{equation}
where $u$ is the envelope of optical pulse, $z$ is the propagation distance and all quantities are
made dimensionless. Consider a two-step periodic dispersion map: $d(z)=\langle d\rangle+\tilde
d(z)$, where $\tilde d(z)=d_1$ for $0<z+n L<L_1$ and $\tilde d(z)=d_2$ for $L_1<z+n L<L$, $L\equiv
L_1+L_2$ is a dispersion map period, $\langle d \rangle$ is the path-averaged dispersion, $d_1,
d_2$ are the amplitudes of dispersion variation subjected to condition $d_1L_1+d_2L_2\equiv 0$ and
$n$ is an arbitrary integer number.

A nonlinearity can be treated as a small perturbation on a scales of dispersion map period $L$
provided a characteristic nonlinear length $Z_{nl}$ of the pulse is large: $Z_{nl}\gg L,$ where
$Z_{nl}=1/|p|^2$ and $p$ is a typical pulse amplitude. Then the eq. $(\ref{nls1})$ is reduced to a
path-averaged Gabitov-Turitsyn \cite{gabtur1} model:
\begin{eqnarray} \label{psiint1}
i \hat \psi_z(\omega) -\omega^2 \langle d \rangle \hat \psi + R(\hat \psi,\omega)  =0,
\end{eqnarray}
where
\begin{eqnarray} \label{Rdef}
R(\hat \psi,\omega)=\frac{1}{(2\pi)^2}\int
\frac{\sin{\frac{s\triangle}{2}}}{\frac{s\triangle}{2} } \hat\psi(\omega_1)\hat\psi(\omega_2)  \nonumber \\
\times\hat\psi^*(\omega_3) \delta(\omega_1+\omega_2-\omega_3-\omega) d\omega_1 d\omega_2 d\omega_3,
\end{eqnarray}
$\triangle\equiv\omega_1^2+\omega_2^2-\omega_3^2-\omega^2$, $s=d_1 L_1$ is a dispersion map
strength, $\hat\psi \equiv \hat u e^{i\omega^2\int^z_{L_1/2} \tilde d(z')dz'}$ is a slow function
of $z$ on a scale $L$ and $\hat\psi(\omega)=\int^\infty_{-\infty}\psi(t)e^{\imath \omega t}dt$ is a
Fourier component of $\psi$. The Gabitov-Turitsyn model is well supported by numerical simulations
\cite{turitsyn1,ablowitz1}.

Consider DM soliton solution $\psi= A(t)e^{i\lambda z} $ ($A$ is real) of the Gabitov-Turitsyn Eq.
$(\ref{psiint1})$ then returning to t-space one gets:
\begin{eqnarray} \label{psiint1c}
-\lambda A + \langle d \rangle A_{tt}= \frac{1}{2\pi s}\int Ci\big(\frac{t_1 t_2}{s}\big )
A(t_1+t)\times   \nonumber
\\ A(t_2+t)A(t_1+t_2+t) dt_1 dt_2,
\end{eqnarray}
where $Ci(x)=\int^x_\infty\cos{x}/xdx$. It was found numerically \cite{smithknox1} that the
Gaussian ansatz
\begin{equation} \label{psi0}
A_{Gauss}=p \exp{\big ( -\frac{\beta}{2}t^2\big )},
\end{equation}
where $p, \beta$ are real constants, is a rather  good approximation for the DM soliton solution.
Thus the Eq. $(\ref{psi0})$ can be effectively used as zero approximation for solving Eq.
$(\ref{psiint1c})$ by iterations which was done in Ref. \cite{lush2000a} for $\langle d \rangle
=0.$ Following \cite{lush2000a} one can easily make a generalization for the case of small but
nonzero average dispersion $|d_0| \ll |d_1|$ and obtain  a set of two transcendental Eqs. from the
series expansion of a first iteration in powers of $t^2$:
\begin{eqnarray} \label{beta2lambda}
   \lambda=-\beta \langle d \rangle + \frac{p^2}{2\sqrt{3}\tilde s}\big(arcsinh
   \frac{3\tilde s-\imath}{2}+c.c.\big),
        \nonumber \\
  \lambda=-3\beta \langle d \rangle
   \nonumber \\
  + \frac{2p^2}{3\tilde s}
  \Big(\sqrt{\frac{\tilde s+\imath}{3\tilde
  s-\imath}}+\frac{\sqrt{3}}{12} arcsinh\frac{3\tilde s-\imath}{2}+c.c. \Big),\nonumber
\end{eqnarray}
where c.c. means complex conjugation. The Eqs. $(\ref{beta2lambda})$ determine the parameters
$\beta, \, p$ of the Gaussian ansatz $(\ref{psi0})$ as a functions of system parameters $\lambda,
\, s$.

The solution of Eq. $(\ref{psiint1c})$ was also obtained by means of iterating this Eq.
numerically. $n+1$th  iteration $A^{(n+1)}$ is given by:
\begin{eqnarray} \label{An1}
\hat  A^{(n+1)}(\omega) = Q_n^{3/2} \frac{R(\hat A^{(n)},\omega)+(|\langle d \rangle|-\langle d
\rangle)\omega^2 \hat A^{(n)}(\omega)}{\lambda+|\langle d \rangle|\omega^2},
\end{eqnarray}
where the functional $R(\hat A,\omega)$ is defined in $(\ref{Rdef})$, $Q_n$ is a stabilizing factor
given by
\begin{eqnarray} \label{Qn}
Q_n=\frac{\hat F^{-1} \Big (\frac{\lambda+\langle d \rangle\omega^2}{\lambda+|\langle d
\rangle|\omega^2}\hat A^{(n)}(\omega)\Big )}{\hat F^{-1} \Big (\frac{R(\hat
A,\omega)}{\lambda+|\langle d \rangle|\omega^2} \Big )}\Big |_{t=0}
\end{eqnarray}
and $\hat F^{-1}$ is a Backward Fourier transform. This numerical iteration scheme was also used in
Ref. \cite{pelin2000} except that a Petviashvili stabilizing factor\cite{petviashvili1992} was
used there instead of $Q_n$. But both stabilizing factors results in the convergence of iteration
scheme to the same solution of $(\ref{psiint1c}).$

The main obstacle in numerical iteration scheme  $(\ref{An1}),(\ref{Qn})$ is the computation of
integral term $R(\hat A,\omega)$ which generally require $N^3$ operations for each iteration,
where $N$ is a number of grid points in $\omega$ or $t$-space. Here we introduce much more
efficient numerical algorithm for calculation of $R(\hat A,\omega)$.

Rewriting the kernel of $R(\hat \psi,\omega)$ via parametric integral:
\begin{equation} \label{paramc1}
\frac{\sin{\frac{s \triangle}{2}}} {\frac{s\triangle}{2} }  = \frac{1}{s}\int_{-s/2}^{s/2}
\exp{\big (i s' \triangle \big)}ds'
\end{equation}
and using definition of $\triangle$ one gets from $(\ref{Rdef}):$
\begin{eqnarray} \label{Rdef2}
R(\hat A,\omega)=\frac{1}{s(2\pi)^2}\int_{-s/2}^{s/2}d s' e^{-i s' \omega^2} \int
\hat A^{(s')}(\omega_1) \nonumber \\
\times\hat A^{(s')}(\omega_2) \hat A^{(s') \, *}(\omega_3)
\delta(\omega_1+\omega_2-\omega_3-\omega) d\omega_1 d\omega_2 d\omega_3,
\end{eqnarray}
where $\hat A^{(s')}(\omega) \equiv \hat A(\omega)e^{is'\omega^2}.$ In $t$-space this expression
takes the form
\begin{eqnarray} \label{Rdef3}
\hat F^{-1}\big (R(\hat A,\omega)\big )=\frac{1}{s}\int_{-s/2}^{s/2}d s'
 {\bf G}^{(s')}\big (\Psi^{(s')}(t)\big),
\end{eqnarray}
where $\Psi^{(s')}(t) \equiv |A^{(s')}(t)|^2A^{(s')}(t)$ and $\bf{ G}^{(s')}$ is an integral
operator corresponding to a multiplication operator $\hat{\bf G}^{(s')}\big
(\hat\Psi^{(s')}(\omega)\big)\equiv e^{is'\omega^2} \hat \Psi^{(s')}$ in $\omega$-space. It follows
from the Eqs. $(\ref{Rdef2}),(\ref{Rdef3})$ that numerical procedure for calculation of $R(\hat
A,\omega)$ includes four steps:

(i) The Backward Fourier Transform of $\hat A^{(s')}(\omega)=\hat A(\omega)e^{is'\omega^2}$ for
every value of $s'.$

(ii) A calculation of $\Psi^{(s')}(t)$ from $A^{(s')}(t)$.

(iii) The Forward Fourier Transform of $\Psi^{(s')}(t).$

(iv) A numerical integration (summation) of $ e^{is'\omega^2}\hat\Psi^{(s')}(\omega)$ over $s'$ for
every value of $\omega.$

The Forward and Backward Fourier Transforms were performed with the Fast Fourier Transform which
requires $N Log_2(N)$ operations. A total number of operations for one iteration is about $4M N
Log_2(N),$ where $M$ is a number of grid points for integration over $s'$. We used the following
typical values for numerical solution of $(\ref{psiint1c})$: $N=8192; \ M=800$. One iteration on
Alpha 500MHz workstation requires about 30 seconds for 16-bytes (32 digits) precision. Thus
numerical scheme (i)-(iv) provides dramatic improvement of numerical performance. $8192^3$
operations would takes 30 days on the same workstation. Note that the proposed efficient numerical
algorithm can be generalized to include optical fiber losses and amplifiers.

Fig.1a,b show the dependence of  a root mean square pulse width $T_{RMS}\equiv \sqrt{\int t^2
A^2dt/\int A^2 dt}$ on a quasimomentum $\lambda$  obtained from $(i)$ the first iteration of the
Eq. $(\ref{psiint1c})$ using values of $\beta, p$ resulting from the Eqs. $(\ref{beta2lambda})$
(dotted curves); $(ii)$ a variational approach (see e.g. Eqs. $(13),(14)$ in Ref. \cite{pelin2000})
represented by dashed lines; $(iii)$  a full numerical solution of the Eq. $(\ref{psiint1c})$
(solid lines).  The explicit expression $T_{RMS}=1/\sqrt{2\beta}$ for the Gaussian pulse shape is
used for calculation of dashed curves.  The solid curve is shown in Fig. 1b only for upper branch I
of solution because numerical iteration scheme for negative average dispersion $\langle d
\rangle=-0.01$ diverges on lower branch II which is in agreement with Ref. \cite{pelin2000}. We
also calculated a time-averaged optical power $P\equiv \int A^2 dt$ and found that $P(\lambda)$
dependence following from the first iteration and the variational approach Ref. \cite{pelin2000}
reproduce a full numerical solution of the Eq. $(\ref{psiint1c})$ with high accuracy ($\sim 1\%$).
One can conclude that both the Eqs. $(\ref{beta2lambda})$ and the variational approach
\cite{pelin2000} predict $P(\lambda)$ with a high accuracy while $T_{RMS}(\lambda)$ dependence is
reproduced by the first iteration of the Eq. $(\ref{psiint1c})$ with better accuracy ($\sim 2\%$)
compare with accuracy ($\sim 40\%$) of the variational approach.

There is an essential difference of our numerical simulation in comparison with numerical results
of Ref. \cite{pelin2000} concerning upper branch I for the negative average dispersion. After about
50 iterations of the Eq. $(\ref{psiint1c})$ a numerical instability was detected on the tails of DM
soliton  for $\langle d \rangle=-0.01$. (Presumably  this instability was not found in Ref.
\cite{pelin2000} because a few iterations was considered there). A finer numerical grid slows down
numerical instability growth but does not kill it. The instability slows down as $\langle d
\rangle\to 0$ and for $\langle d \rangle\ge 0$ there is no numerical instability.  Thus the solid
curve in Fig. 1b for $\langle d \rangle=-0.01$ can only be formally attributed to DM soliton and
the question about existence of DM soliton for the negative average dispersion is still open. It is
possible that the instability within the numerical iteration scheme results from a resonance of DM
soliton tails with linear waves \cite{pelin2000}. But there is another alternative that DM soliton
solution does not exist for any negative average dispersion value and instead of DM soliton on can
observe a long-lived quasi-stable structure. Note that the existence of DM soliton for nonnegative
average dispersion for the Eq. $(\ref{psiint1c})$  was proved in Ref. \cite{zharnitsky2000a}. In
addition it was proved in Ref. \cite{zharnitsky2000b} that even if DM soliton exists for $\langle
d \rangle<0$ it can not realize a minimum of the Hamiltonian of the Eq.  $(\ref{psiint1})$ for
fixed $P$. This can indicate that DM soliton is unstable in that case. Related result
\cite{lush2000b} is the nonexistence criterion for a periodic solution of the Eq. $(\ref{nls1})$
for a negative enough average dispersion. But Refs.
\cite{zharnitsky2000a,zharnitsky2000b,lush2000b} do not give any statement about the existence of
DM soliton for small negative $\langle d\rangle$.

Fig. 2 shows a typical shape of DM soliton. This is the first to the best of my knowledge high
precision numerical solution of the Eq. $(\ref{psiint1c})$. Note that solid curve dips do not
reach $t-$axes only because of finite size of numerical grid. On can conjecture from Fig. 2 the
asymptotic of DM soliton is given by
\begin{equation} \label{psishort1}
A_{asymp}(t)=f(t)\cos{\Big (t^2\big [a_0+a(t)\big]\Big)} \exp{(-b |t|)},
\end{equation}
where $a_0, b $ are constants and $f(t)/|t|, \, a(t)$ are slow functions of $t.$ An analysis of
fast oscillations in integral term of the Eq.  $(\ref{psiint1c})$ allows to show that
$f(t)=c|t|+O(1), \ a_0=1/2 s, \, a(t)=a_1 /|t|+a_2/t^2+O(1/|t|^3)$ for $|t|\to \infty, \ \langle
d\rangle\to 0$, where $c, a_1, a_2$ are constants. Dashed curve in Fig. 2 shows $A_{asymp}^2(t)$
dependence for $c=11.9654,\, b=3.04515,\, a_1=1.41364,\, a_2=1.51023$ which is in very good
agreement with asymptotic of numerical solution of the Eq. $(\ref{psiint1c})$ (solid curve). Thus
the envelope of DM soliton oscillating tails decays exponentially  while the oscillations are
described by a quadratic law. Detailed consideration of the asymptotic solution is outside the
scope of this Letter.

The author thanks M. Chertkov, I.R. Gabitov, E.A. Kuznetsov and V. Zharnitsky for helpful discussions.

The support was provided by  the Department of Energy, under contract W-7405-ENG-36, RFBR (grant
00-01-00929) and the program of government support for leading scientific schools (grant
00-15-96007).

E-mail address:  lushnikov@cnls.lanl.gov

\newpage

\section*{Figure captions:}

~

\noindent Fig.1. $T_{RMS}$ for $s=1$,$\langle d \rangle =0.01$ (a) and $\langle d \rangle =-0.01$
(b).  Branches I, II for $\langle d \rangle =-0.01$ correspond to two branches of analytical
solution.

\noindent Fig.2. DM soliton shape (curve 1) versus Eq. (11) (curve 2) for $\langle d\rangle=0,\
s=1, \ \lambda=1.$ $A(t)$ is an even function.


\begin{figure*}
\epsfxsize=10.5cm \epsffile{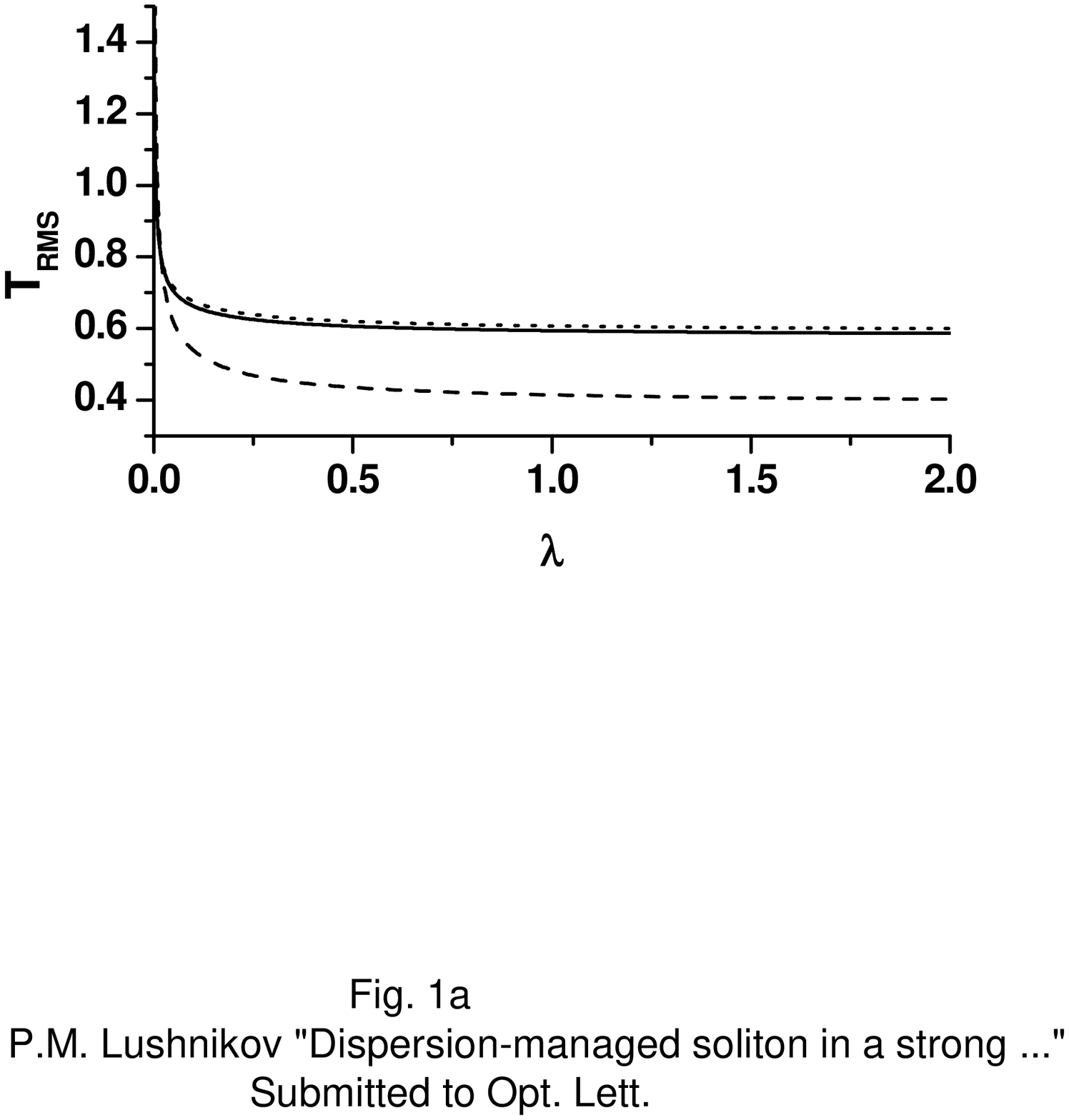}
\end{figure*}

\begin{figure*}
\epsfxsize=10.5cm \epsffile{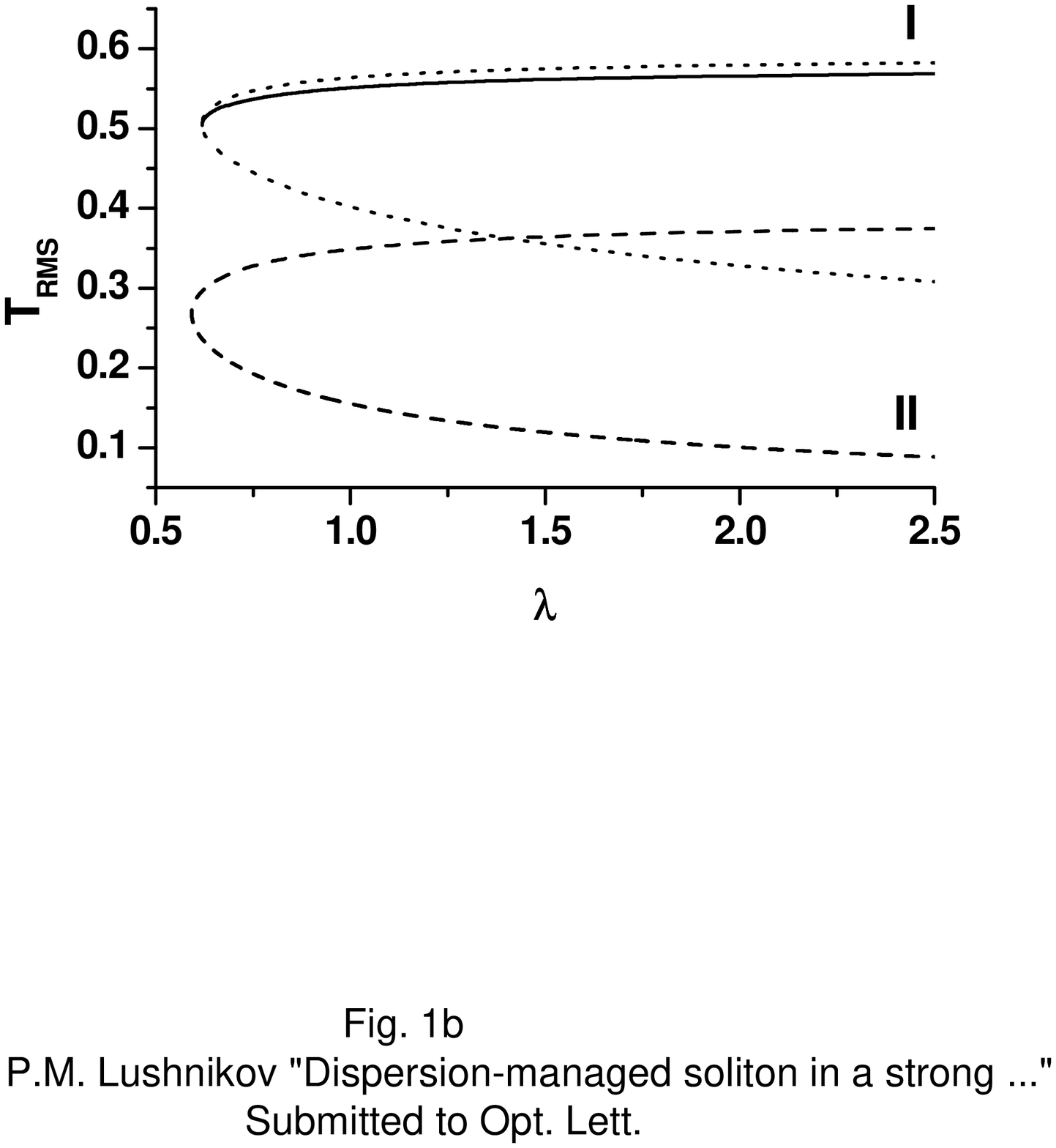}
\end{figure*}

\begin{figure*}
\epsfxsize=15.5cm \epsffile{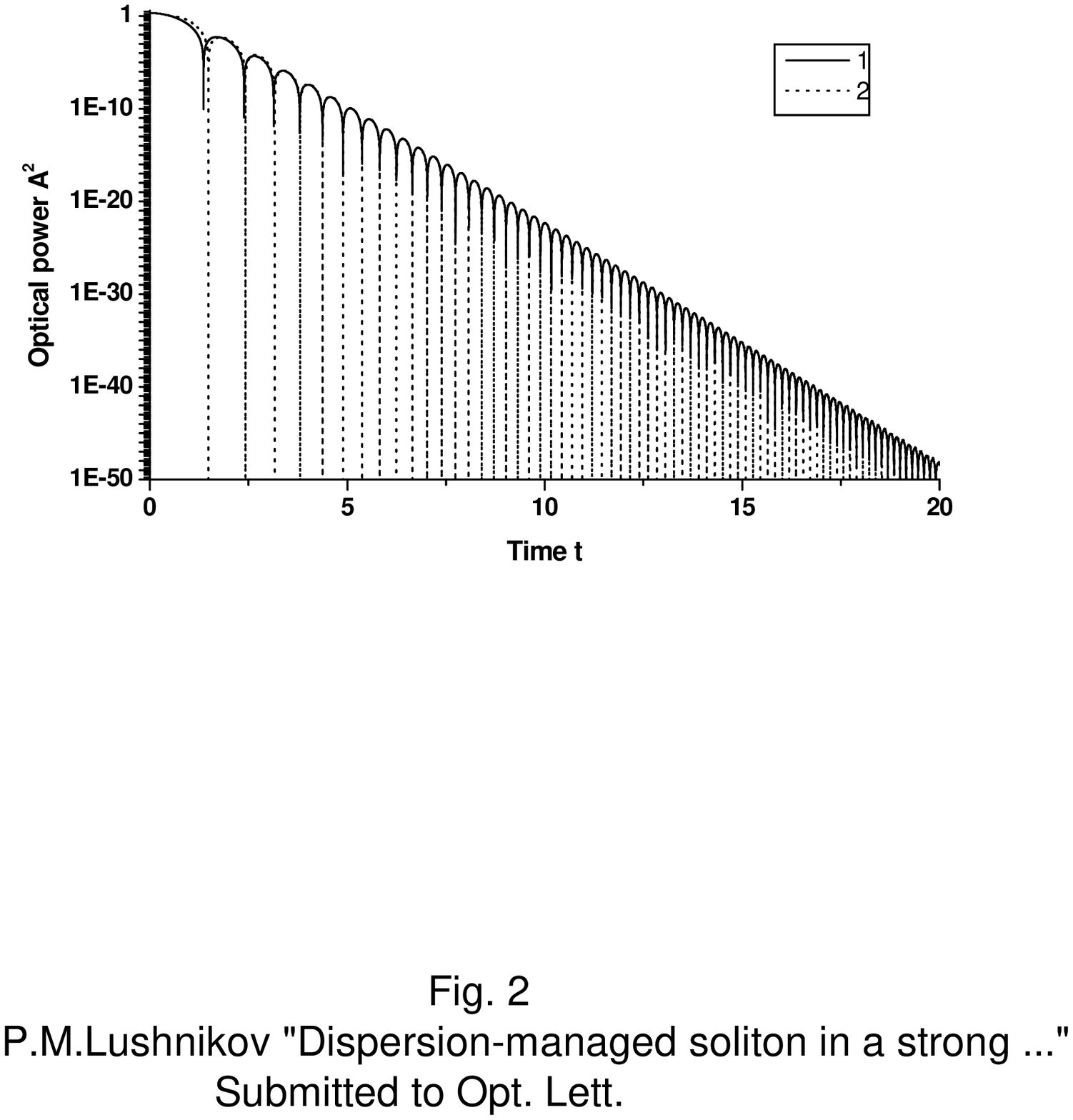}
\end{figure*}

\end{document}